\documentclass[aps,twocolumn,showpacs]{revtex4}
\usepackage{graphicx}

\def\ss{\scriptscriptstyle }
\begin{document}
\title{Thermal-radiation-induced nonequilibrium carriers in an intrinsic graphene}
\author{P.N. Romanets}
\author{F.T. Vasko}
\email{ftvasko@yahoo.com}
\author{M.V. Strikha}
\affiliation{Institute of Semiconductor Physics, NAS of Ukraine,
Pr. Nauky 41, Kyiv, 03028, Ukraine}
\date{\today}

\begin{abstract}
We examine an intrinsic graphene connected to the phonon thermostat at temperature
$T$ under irradiation of thermal photons with temperature $T_r$, other than $T$. The
distribution of nonequilibrium electron-hole pairs was obtained for the cases of
low and high concentration of carriers. For the case when the interparticle scattering
is unessential, the distribution function is determined by the interplay of intraband 
relaxation of energy due to acoustic phonons and interband radiative transitions caused 
by the thermal radiation. When the Coulomb scattering dominates, then the
quasi-equilibrium distribution with effective temperature and non-equilibrium
concentration, determined through balance equations, is realized. Due to the effect
of thermal radiation with temperature $T_r\neq T$ concentration and conductivity
of carriers in graphene modify essentially. It is demonstrated, that at $T_r>T$
the negative interband absorption, caused by the inversion of carriers distribution,
can occur, i.e. graphene can be unstable under thermal irradiation.
\end{abstract}


\maketitle
Different kinetic phenomena caused by carriers localized near the band cross-point of
graphene, including dc (magneto)transport and optical properties, have been studied
intensively within recent years, see reviews \cite{1} and last references in
\cite{2,3}. The main attention was paid to examination of linear response of the
carriers in the phonon thermostat at the temperature $T$. Because of the weak carrier
interaction with acoustic phonons \cite{4} different external factors can easily
disturb the equilibrium of electron-hole system, and the linear responce behaviour
realization needs accurate control. In particular, when the sample is not isolated
from external thermal radiation with temperature $T_r\neq T$, the carriers interaction
with additional thermostat of thermal photons is essential. This interaction can
be effective enough, because the interband transitions are determined by the
velocity $v_W=10^8$ cm/s characterizing the linear spectrum of carriers (the
neutrino-like states near the band-crossing point are described by the Weyl-Wallace
model \cite{5}). Therefore graphene is very sensitive for thermal irradiation: the
concentration and conductivity of carriers modify essentially (particularly, the 
photoconductivity induced by thermal irradiation occurs, compare with \cite{6}, 
where the case of the interband pumping was discussed).

In this paper, the results for the distribution of non-equilibrium carriers 
in the intrinsic graphene, interacting with phonon and proton thermostats with different temperatures, are presented. This distribution is obtained from the kinetic equation, 
taking into consideration the quasi-elastic energy relaxation due to acoustic phonons, 
and generation-recombination processes due to interband transitions, caused by thermal irradiation (the corresponding collision integrals were obtained in \cite{6}). Under 
the high concentrations it is also necessary to take into consideration the Coulomb 
scattering, which does not cause the interband transitions, see \cite{7}. Moreover, the scattering by static disorder should be taken into account as the main mechanism of momentum relaxation \cite{8}. In the low temperatures range, where the carrier concentration is 
not high, the Coulomb scattering is unessential and the distribution function differs 
essentially from equilibrium one due to interplay between acoustic scattering and radiative transitions. At high temperatures, when the Coulomb scattering dominates, the quasi-equilibrium
distribution of carriers, with effective temperature and non-equilibrium concentration,
is imposed. We also calculate the dependences of concentration and conductivity on
$T$ and $T_r$ under the scattering by the short-range static disorder.

In the intrinsic graphene with the symmetrical $c$- and $v$-bands, and with similar
scattering in these bands, the distributions of electrons and holes are identical;
they are described by distribution function $f_{p}$. This function is governed by the
quasi-classic kinetic equation \cite{6}:
\begin{equation}
J_{\ss LA}(f|p)+J_{\ss R}(f|p)+J_{\ss C}(f|p)=0 .
\end{equation}
Here the collision integrals $J_{\ss LA}$, $J_{\ss R}$ and $J_{\ss C}$ describe the
relaxation of carriers caused by the phonon ($LA$) and photon ($R$) thermostats
and the carrier-carrier scattering ($C$), respectively. The solutions of Eq (1) have
been obtained below for the two cases: $({\it a})$ low concentrations, when $J_{\ss C}$
can be neglected, and $({\it b})$ high concentrations, when $J_{\ss C}$ imposes the
quasi-equilibrium distribution with parameters determined from the equations of the 
balance of concentration and energy. After summation of Eq. (1) over $\bf p$-plane with 
the weights 1 and $p$ we get these balance equations in the form \cite{9}:
\begin{eqnarray}
\frac{4}{L^2}\sum_{\bf p}J_{\ss R}( f_t|p)=0,  \\
\frac{4}{L^2}\sum_{\bf p}p[J_{\ss LA}( f_t|p)+J_{\ss R}( f_t|p)]=0 ,
\end{eqnarray}
where the contribution of acoustic scattering is omitted from the equation of the
concentration balance, because the interband transitions are forbidden due to the
inequality $s\ll v_W$. The inter-carrier scattering does not change the concentration
and energy. \cite{7}

We start with the examination of low temperature case, when the carriers concentration 
is small and $J_{\ss C}$ in Eq.(1) can be neglected. With the use of the collision integrals
$J_{\ss LA}$ and $J_{\ss R}$, presented in \cite{6}, with temperatures $T$ and $T_r$
correspondingly, we get the non-linear equation of the second order for the distribution
function $f_{p}$:
\begin{eqnarray}
\frac{\nu_p^{\ss (qe)}}{p^2}\frac{d}{dp}\left\{ p^4\left[\frac{df_p}
{dp}+\frac{f_p (1 -f_p )}{p_{\ss T}}\right]\right\} \nonumber  \\
+ \nu_p ^{\ss (r)}\left[ N_{2p/p_r}(1-2f_{pt})-f_{pt}^2\right] =0 .
\end{eqnarray}
Here we introduce the characteristic momenta $p_T=T/v_W$ and $p_r=T_r/v_W$;
$N_{2p/p_r}=[\exp (2p/p_r)-1]^{-1}$ is the Plank function. The rates of quasi-elastic 
relaxation at acoustic phonons, $\nu_p^{\ss (qe)}=(s/v_W)^2v_{ac}p /\hbar$, and radiative transitions, $\nu_p^{\ss (r)}=v_rp/\hbar$, have been expressed through the sound velocity, 
$s$, and through the characteristic velocities, $v_{ac}$, and $v_r$. They separate the
momentum dependence of the relaxation rates, which is proportional to the density of 
states. According to \cite{8}, where the temperature dependence of mobility have been 
examined, $v_{ac}\simeq 1.35\cdot 10^4$ cm/s for the helium temperature (moreover 
$v_{ac}\propto T$), and the quasi-elastic character of scattering is determined by the 
small parameter $s/v_W$. For the case of graphene, placed between the SiO$_2$ substrate 
and cover layer, we get $v_r\simeq$41.6 cm/s, see \cite{6}.

The boundary conditions in Eq.(4) are imposed both by the demand of the finite stream 
along the energy axis, when $p^4(df_p/dp+f_p)_{p\to\infty}<const$, and by the
concentration balance equation (2). Because the acoustic and radiative contributions become 
zero under the equilibrium contributions with temperatures $T$ and $T_r$ correspondingly, 
and these two contributions trend to 1/2 at small $p$, we get $f_{p\rightarrow 0}=1/2$. 
\cite{10} This demand can be used as the boundary condition for the numerical solution 
of Eq.(4). Note, that the validity of Eq. (2) for the obtained distribution should be examined. This solution was carried out below with the use of the finite difference method 
and the iterations over non-linear contributions in Eq.(4), see \cite{11}.

\begin{figure}[ht]
\begin{center}
\includegraphics{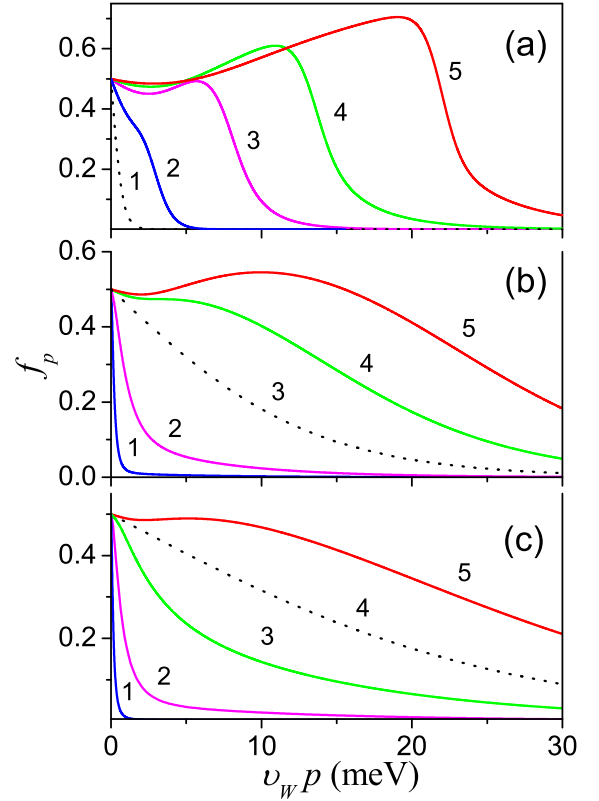}
\end{center}
\addvspace{-1 cm}
\caption{(Color online) Distribution functions governed by Eq. (4) for the cases: 
$T$=4.2 K (a), $T$=77 K (b), and $T$=150 K (c) at $T_r$=4.2 K (1), 20 K (2), 77 K (3), 
150 K (4), and 250 K (5). Dotted curves are the equilibrium thermal distributions at
$T=T_r$.}
\end{figure}
In Fig. 1a-c the obtained distribution functions versus energy $v_Wp$ for different $T$ 
and $T_r$ are presented. Due to the smallness of acoustic contribution at $p\to 0$ (see 
above), $f_p$ is close to equilibrium distribution with the temperature $T_r$. If $T_r>T$ 
the distribution increases at high energies up to the range, where $J_{LA}$ is dominant. 
With the further increase of $v_Wp$ the distribution decreases rapidly on the scale of 
energies $T$, so that the peak of distribution is formed at the energies $\sim T_r$; 
this peak causes the carriers concentration of the order of equilibrium value at 
the temperature $T_r$. In this case the condition $f_{max}>1/2$ is realised, i.e. the 
inverse distribution of the carriers occupation takes place for the energies close to 
the maximum of distribution (see below). On the contrary, if $T_r<T$, the equilibrium contribution of the slow carriers is replaced by the rapidly decreasing part in the range
of $v_Wp<T$. This distribution determines the small concentration of carriers 
in comparison with the equilibrium value at the temperature $T$.

\begin{figure}[ht]
\begin{center}
\includegraphics{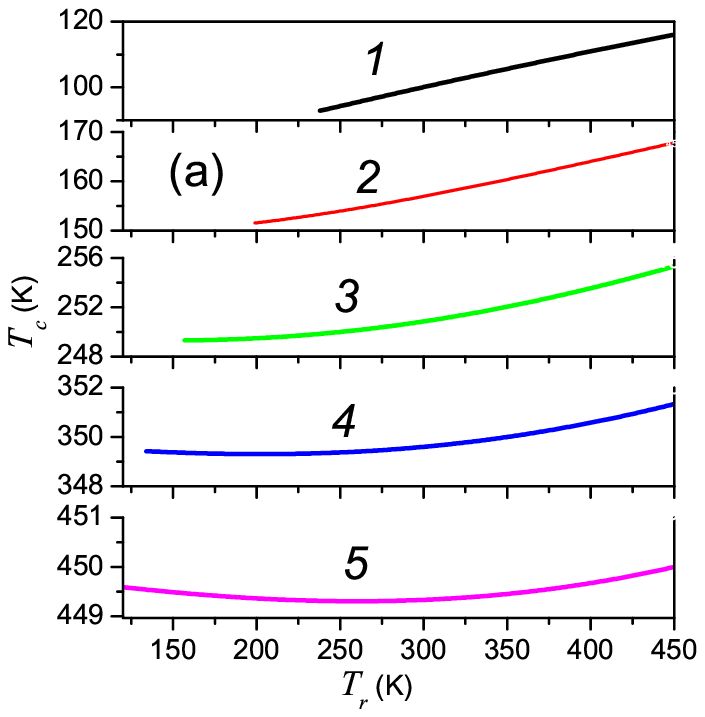}
\includegraphics[scale=2]{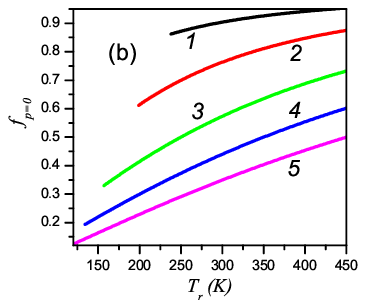}
\end{center}
\addvspace{-1 cm}
\caption{(Color online) Effective temperature $T_c$ (a) and maximal distribution
$\tilde{f}_{p=0}$ (b) versus $T_r$ for different temperatures $T=$77 K (1), 
150 K (2), 250 K (3), 350 K (4) and 450 K (5).}
\end{figure}
Later we shall examine the case of high temperatures (and concentrations), when the Coulomb scattering dominates, imposing the quasi-equilibrium distribution
\begin{equation}
\tilde{f}_p=\{\exp [(v_Wp-\mu )/T_c]+1\}^{-1} .
\end{equation}
The effective temperature of carriers $T_c$ and the chemical potential $\mu$ in this distribution are obtained from the balance equations (2) and (3). After introducing the dimensionless momentum $x=v_Wp/T_c$, we get the equation of the concentration balance
\begin{equation}
\int_0^{\infty}dxx^2\left(\frac{1-2\tilde{f}_x}{e^{2xT_c/T_r}-1}
-\tilde{f}_x^2\right) =0 ,
\end{equation}
which imposes the relation between $\mu$, and $T_c/T_r$, while the phonon temperature 
is omitted out of this equation. In these variables the equation of the energy balance 
can be presented as:
\begin{eqnarray}
\int_0^{\infty}dxx^3\left(\frac{1-2\tilde{f}_x}{e^{2xT_c/T_r}-1}-\tilde{f}_x^2
\right) \nonumber  \\
-\gamma\frac{T_c-T}{T}\int_0^{\infty}dxx^4e^{x-\mu /T_c}\tilde{f}_x^2 =0 , ~~~
\end{eqnarray}
where $\gamma =(s/v_W)^2v_{ac}/v_r\propto T$ determines the relative contribution of the 
phonon and photon thermostats.

The solution of the transcendental equations (6) and (7) gives the distribution (5), 
dependent on $T$ and $T_r$, which can be characterised by the effective temperature $T_c$, 
and the maximum value of the function $\tilde{f}_{p=0}=[\exp (-\mu /T_c)+1]^{-1}$.
These values are presented in Fig.2 for the concentrations, greater than $3.5\cdot 10^{10}$ 
cm$^{-2}$. Note, that the temperature $T_c$ differs from $T$ unessentially under the great 
change of $T_r$, meanwhile $\mu$ and $\tilde{f}_p$ modify essentially for the temperature 
range up to 450 K (the high-temperature measurements of graphene were carried out recently 
\cite{12}). The essential difference of the case under examination from the low temperature solution, presented in Fig.1, is that the value $f_{p=0}$ is being fixed (and equal to 1/2), 
while the value $\tilde{f}_{p=0}$ decreases with the increase of $T$, and increases with 
the increase of $T_r$; note, that $\tilde{f}_{p=0}$=1/2 when $T=T_r$.

\begin{figure}[ht]
\begin{center}
\includegraphics{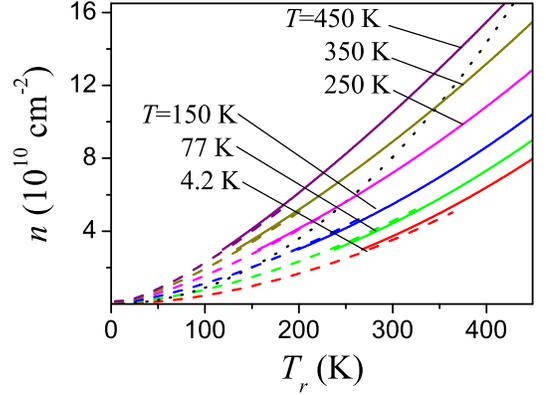}
\end{center}
\addvspace{-1 cm}
\caption{(Color online) Carrier concentrations given by Eq. (8) versus temperature $T_r$ 
for different $T$. Solutions of Eq. (4) are plotted in the range $n<5\cdot 10^{10}$
cm$^{-2}$ (dashed curves). Solutions of the balance equations (6) and (7) are plotted
in the range $n>3.5\cdot 10^{10}$ cm$^{-2}$ (solid curves). The equilibrium
concentration (if $T=T_r$) is shown as dotted curve.}
\end{figure}
As one can see from Figs. 1a,b or 2b, the distribution $f_p$ or $\tilde{f}_p$ can be greater than 1/2 in a certain energy range or at low energies. Since the inversion of 
electron-hole pairs occupation, the regime of the negative interband absorption 
can occur because the real part of dynamic conductivity is given by expression \cite{3} $Re\sigma_\omega =(e^2/4\hbar )(1-2f_{p_\omega})$. In the range of parameters under 
examination the intrinsic graphene tends to be unstable, if an additional adsorption 
is weak enough.

The distributions, presented in Figs. 1 and 2, are valid for the cases of low and high concentrations, correspondingly. For the calculation of non-equilibrium concentration $n$, dependent on $T$, and $T_r$, we use the standard expression
\begin{equation}
n=\frac{2}{\pi\hbar^2}\int\limits_0^\infty dppf_{p}=\frac{2}{\pi}\left(\frac{T_c}
{\hbar v_W}\right)^2\int_0^\infty dxx \tilde{f}_x ,
\end{equation}
where the right equality was written under the substitution of Eq.(5) into the standard 
formula. In Fig. 3 we plot $n$ versus $T$ and $T_r$ for the low- and high temperature 
regions, when the Coulomb scattering can either be neglected, or it dominates. At $T=T_r$ 
these curves intersect with equilibrium concentration $\propto T^2$: because the 
concentration is controlled by thermal irradiation, $n$ is smaller (or greater) then 
equilibrium concentration at $T<T_r$ (or $T>T_r$). Note, that the dependences obtained 
from equation (4), and from the equations of balance (6), (7) correspond well in the 
range of intermediate concentrations, $3.5\div 5\cdot 10^{10}$ cm$^{-2}$.

\begin{figure}[ht]
\begin{center}
\includegraphics{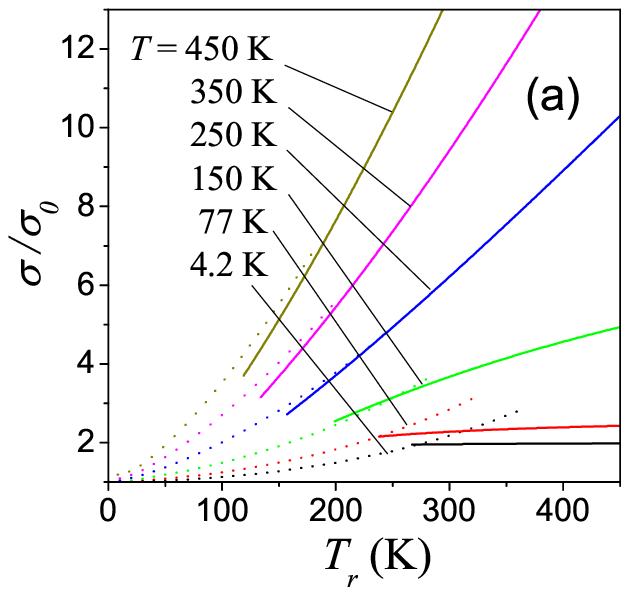}
\includegraphics{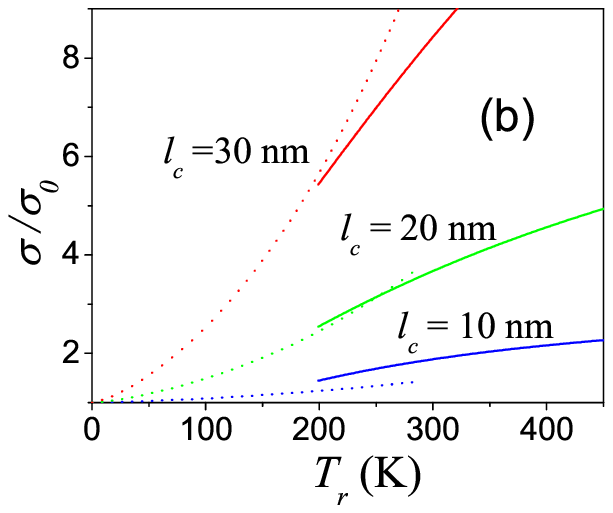}
\end{center}
\addvspace{-1 cm}
\caption{(Color online) (a) Normalized conductivity, $\sigma /\sigma_o$, versus $T_r$ for different $T$ at $l_c$=20 nm. (b) The same for $T$=150 K at different correlation 
lengths, $l_c$.}
\end{figure}

The modifications of the distribution of non-equilibrium carriers in the intrinsic graphene, described above, lead to modification of conductivity, $\sigma$. For the scattering of momenta
due to static disorder with correlation length $l_c$ we calculate the conductivity, 
using the formula \cite{9}
\begin{equation}
\sigma = \sigma_o\biggl[2f_{p=0}-\frac{l_c}{\hbar}
\int_0^{\infty}dpf_p\frac{\Psi '(pl_c/\hbar)}{\Psi (pl_c/\hbar)^2}\biggr] ,
\end{equation}
where $\Psi (z)=\exp (-z^2)I_1(z^2)/z^2$ is written through the first order Bessel function 
of the imagined argument, $I_1(z)$, and $\sigma_o$ is the conductivity in the case of 
short-range disorder scattering, when $l_c = 0$. For the case of short-range scattering, $\overline{p}l_c/\hbar\ll 1$ ($\overline{p}$ is the characteristic momentum of non-equilibrium carriers) the conductivity is written through the low-temperature distribution: $\sigma 
\simeq 2\sigma_of_{p=0}$. For the low temperature range, when $f_{p=0}=1/2$, the conductivity depends weakly on $T$, and $T_r$. On the contrary, in the high temperature range the 
temperature dependences of $\sigma$ on $T$ and $T_r$ are essential, see Fig. 4a, and Fig. 
2b, where $\tilde{f}_{p=0}$ is plotted. With the increase of $l_c$ the second term in 
Eq. (9) becomes essential and the thermal dependences $\sigma /\sigma_o$ become much 
stronger, see Fig.4b.

Next, we list the assumptions used. The main restriction is the examination of the limit 
cases either of no inter-carrier collisions, or of their domination only. Despite some 
results correspond well in the intermediate range of concentrations and temperatures
(a disagreement of $\sigma /\sigma_o$, versus $T_r$ takes place at $l_c<$20 nm only, see
Fig. 4b), the accurate analysis of the intermediate range is beyond of the scope of this 
paper. Other assumptions, such as the models of energy spectrum, or the scattering mechanisms, 
as well as the simplifications of distribution function used, are rather 
standard for the calculations of the transport phenomena.

In closing, the consideration performed demonstrates the essential effect of thermal 
irradiation on the intrinsic graphene properties, therefore the transport measurements 
should be carried under the control of the condition $T=T_r$. Within the study of the 
device applications we should also consider the possible difference between the temperature 
of phonons and the temperature of the external thermal radiation.


\begin{thebibliography}{52}
\bibitem{4}
The optic phonons are freezed out at $T\ll$ 0.2 eV, therefore they are unessential even for 
the temperatures, higher then the room one.
\bibitem{10}
In order to check the condition $f_{p\rightarrow 0}=1/2$ we have examined the transient
process of the formation of the steady-state distribution, described by the equation
$$
\frac{\partial f_{pt}}{\partial t}=J_{\ss LA}(f_t|p)+J_{\ss R}(f_t|p)
$$
with the initial condition $t_{pt=0}=[\exp (p/p_T)+1]^{-1}$. The numerical computation of 
the Cauchy problem with the use of time iterations method gives the same steady-state 
distribution, as in Fig. 1.
\end{thebibliography}
\end{document}